\documentclass[aps,pra,showpacs,amsmath,amssymb,amsfonts,lengthcheck,longbibliography,superscriptaddress]{revtex4-2}

\usepackage{graphicx}
\usepackage[caption=false]{subfig}
\usepackage[colorlinks=true,citecolor=blue,linkcolor=blue,urlcolor=blue]{hyperref}

\usepackage{soul}
\usepackage[utf8]{inputenc}
\usepackage[T1]{fontenc}

\usepackage{mathtools}
\usepackage{SIunits}
\usepackage{braket}

\begin{document}

\title{Excess work in counterdiabatic driving}

\author{Lucas P. Kamizaki}
\email{kamizaki@ifi.unicamp.br}
\affiliation{Instituto de F\'isica Gleb Wataghin, Universidade Estadual de Campinas, 13083-859, Campinas, S\~{a}o Paulo, Brazil}

\author{Marcus V. S. Bonan\c{c}a}
\email{mbonanca@ifi.unicamp.br}
\affiliation{Instituto de F\'isica Gleb Wataghin, Universidade Estadual de Campinas, 13083-859, Campinas, S\~{a}o Paulo, Brazil}

\begin{abstract}
Many years have passed since the conception of the quintessential method of shortcut to adiabaticity known as counterdiabatic driving (or transitionless quantum driving). Yet, this method appears to be energetically cost-free and thus continually challenges the task of quantifying the amount of energy it demands to be accomplished. This paper proposes that the energy cost of controlling a closed quantum system using the counterdiabatic method can also be assessed using the instantaneous excess work during the process and related quantities, as the time-averaged excess work. Starting from the Mandelstam-Tamm bound for driven dynamics, we have shown that the speed-up of counterdiabatic driving is linked with the spreading of energy between the eigenstates of the total Hamiltonian, which is necessarily accompanied by transitions between these eigenstates. Nonetheless, although excess work can be used to quantify energetically these transitions, it is well known that the excess work is zero throughout the entire process under counterdiabatic driving. To recover the excess work as an energetic cost quantifier for counterdiabatic driving, we will propose a different interpretation of the parameters of the counterdiabatic Hamiltonian, leading to an excess work different from zero. We have illustrated our findings with the Landau-Zener model.   
\end{abstract} 

\maketitle

\section{Introduction}
\label{sec: Introduction}

The crucial role that quantum physics has been playing in technological development has been recognized, for instance, by the United Nations declaring 2025 the Year of Quantum Science and Technology. The viability of quantum technologies often relies on optimizing, speeding, and increasing the control of quantum processes. Regarding these goals, one class of methods that are fundamental in enhancing the performance of quantum processes is shortcuts to adiabaticity (STA): methods that try to achieve adiabatic evolution at finite time. In other words, STA major goal is to speed up quantum processes while maintaining the evolution as close as possible to the adiabatic evolution.  Here, the term adiabatic is a synonym for quasistatic, since we are considering that the quantum system is not in contact with a thermal bath. In addition to shortcuts to adiabaticity saving time, they also appear to be energetically cost-free. Counterdiabatic driving \cite{demirplak2003adiabatic, demirplak2005assisted, demirplak2008consistency, berry2009transitionless}, a paradigmatic example of a STA, makes this free launch more remarkable. This method relies upon adding an extra control Hamiltonian that suppresses the usual transitions due to the finite-time control and keeps the original eigenstates in their adiabatic evolution as if the process were quasistatic.One might think that this extra control leads to an increase in the energetic cost, but this improvement in performance does not appear to come with a price measured by the average work performed by the control agent \cite{campbell2017trade, funo2017universal}. Several papers have addressed this issue by proposing different quantifiers for the energetic cost. We summarize some of these contributions in what follows:

\begin{itemize}

\item Chen and Muga \cite{chen2010transient} made an initial attempt to quantify the energy cost in the compression/expansion of a quantum harmonic oscillator by using the mean energy and the time-averaged mean energy. Although they focus on a "bang-bang" method \cite{salamon2009maximum} of shortcut to adiabaticity, they also have analyzed the mean energy of a process under counterdiabatic driving.
    
\item Santos and Sarandy \cite{santos2015superadiabatic, santos2016shortcut} tried to quantify the energetic cost of superdiabatic computation using the time-integrated Hilbert-Schmidt norm of the total Hamiltonian subjected to counterdiabatic driving. The idea is that this norm contains information about the spectrum and strength of the Hamiltonian. 

\item Zheng \textit{et al.} \cite{zheng2016cost} proposed a cost measure similar to that of Santos and Sarandy, quantifying the power to generate external driving through the Frobenius norm of the counterdiabatic Hamiltonian. Later, Campbell and Deffner \cite{campbell2017trade} unraveled a trade-off relation between the speed of the quantum driving and this cost. 

\item Funo \textit{et al.} \cite{funo2017universal} have shown, using the two-point energy measurement scheme, that the average work from counterdiabatic driving is equal to the adiabatic work throughout all evolution. However, the variance of the work changes from the variance in the adiabatic case, and this difference integrated during the protocol is used as an energetic cost quantifier. 

\item Kiely \textit{et al.} \cite{kiely2022classical} considered the contribution to the energetic cost of the classical control apparatus. By classically modeling the control, they show the entropy production is linked with Joule heating and Johnson-Nyquist fluctuations.  They also pointed out the similarity between entropy production and the energetic cost proposed by Zheng \textit{et al.}. 

\item Calzetta \cite{calzetta2018not} has shown the deviation of the quantum system from the adiabatic trajectory if we consider the quantum nature of the control device, and thus the cost can be measured by conventional work definitions. 

\item In the context of quantum thermal machines and using a shortcut to the adiabaticity method of dynamical invariants, del Campo \cite{campo2014more} has shown the difference between work and adiabatic work during the process and uses the dissipated work averaged over time as the driving cost. 

\end{itemize}

In all these proposals, the term cost does not have a well-defined meaning and is generally used without sustaining its interpretation as energy consumption, as pointed out in reference \cite{guery2019shortcuts}.  Generally, these cost measures quantify different aspects of the energetics of the system. In this paper, we illustrate the connection between the Mandelstam-Tamm bound for driven dynamics and excess work and how the latter can be applied to counterdiabatic driving.  

We begin in Sec.~\ref{sec: Preliminaries} by introducing some basic concepts, such as the adiabatic theorem, the excess work, and counterdiabatic driving. Then, in Sec.\ref{sec: MLbound}, we show that the spreading of energy in the eigenstates of the Hamiltonian decreases the quantum speed limit imposed by the Mandelstam-Tamm bound and the relation between the energy spreading and the excess work. 

We highlight that the relation between quantum speed limits and work is part of the puzzle of understanding the thermodynamics of quantum systems. While thermodynamics involves subjects as the maximal power and efficiency that quantum thermal machines can achieve, quantum mechanics gives additional constraints to be considered, in the form of bounds to the speed of quantum evolution.

Thus, we discuss the cost of counterdiabatic driving in terms of excess work, while keeping in mind its relation with the quantum speed limits. However, application of the excess work to counterdiabatic driving is not straightforward, so in Sec.\ref{sec: time-averaged_work_in_cd}, we show the well-known failure of applying the excess work to quantify the cost of a process under counterdiabatic driving. Then, in Sec. \ref{sec: AdiabaticLimit}, we show how to reintroduce the excess work as an energetic cost quantifier, giving the Landau-Zener model as an example.   Lastly, in Section \ref{sec: Discussion}, we discuss the thermodynamic expected characteristics of an energetic cost. 

Note that a relation between the quantum speed limit and fluctuations of the excess work was already discussed in Ref\cite{funo2017universal}. Here, we differ by proposing that the excess work itself is an energetic quantifier in counterdiabatic driving. 

\section{Preliminaries}
\label{sec: Preliminaries}
\subsection{Thermally isolated quantum systems and the adiabatic theorem}
The state $|\psi(t) \rangle $ of a thermally isolated quantum system subjected to a time-dependent Hamiltonian $H(t)$ has its time evolution given by Schrödinger equation. Considering the parametrization $s = (t-t_{i})/(t_{f}-t_{i}) = (t-t_{i})/\tau$, where $t_{i}$ and $t_f$ are initial and final time instants, respectively, Schrödinger equation can be written as
\begin{equation}
i \hbar \frac{1}{\tau}\frac{d }{ds}|\psi(s) \rangle = H(s)|\psi (s) \rangle, 
\label{eq: Schrodinger equation}
\end{equation}
with $\tau = t_{f}-t_{i}$ known as protocol time and $s$ varying from $0$ to $1$. The instantaneous eigenvalues $E_{n}(s)$ and eigenstates $|n(s)\rangle$ of $H(s)$ are given by 
\begin{equation}
H(s) |n(s) \rangle = E_{n}(s) |n(s) \rangle. 
\label{eq: eigen equation}
\end{equation}
In this paper, we will only consider nondegenerate Hamiltonians, that is, $E_{m}(s) \neq E_{n}(s)$ for $m \neq n$ for any $s$. The solution of Eq.(\ref{eq: Schrodinger equation}) can be written as
\begin{equation}
|\psi(s) \rangle = \exp_{>}\left[\frac{\tau}{i \hbar} \int_{0}^{s}H(s')ds'\right] |\psi(0)\rangle  ,
\end{equation}
where $\exp_{>}$ is the time-ordered exponential. Although we can write this formal solution, its computation even for the simplest cases can be very cumbersome. However, a much simpler solution is provided by the adiabatic theorem when the Hamiltonian is slowly varying \cite{bussey2021modern}, i.e., when $\tau \rightarrow \infty$. It states that if the system is initially in the eigenstate $|n(0) \rangle$ of the Hamiltonian $H(0)$, it will remain in the corresponding eigenstate $|n(s) \rangle$ of $H(s)$ during the evolution, differing only by a global phase factor. Mathematically, we can express this prediction as
\begin{equation}
|\psi(s) \rangle_{\text{ad}} = e^{i\gamma_{n}(s)}e^{-i\omega_{n}(s)}|n(s)\rangle,
\label{eq: adiabatic theorem}
\end{equation}
where 
\begin{equation}
\gamma_{n}(s) = i\int_{0}^{s} \langle n(s') | \frac{d}{ds'} |n (s') \rangle ds'
\label{eq: geometric phase}
\end{equation}
and
\begin{equation}
\omega_{n}(s) = \frac{\tau}{\hbar}\int_{0}^{s} E_{n}(s')ds'
\label{eq: dynamical phase}
\end{equation}
are the geometric and dynamic phases, respectively. Therefore, in an adiabatic evolution, there are no transitions between different eigenstates of the Hamiltonian.  

\subsection{Mean energy and work}
The mean energy $\mathcal{E}(s)$ is defined by
\begin{equation}
\mathcal{E}(s) \equiv \text{Tr}\{\rho(s)H(s)\},
\label{eq: mean energy}
\end{equation}
where $\rho(s)$ represents the density matrix that, for a pure state, is equal to $\rho(s) = | \psi(s) \rangle \langle \psi(s)|$. During a time-dependent driving, the Hamiltonian and the state change over time, leading to a variation in the mean energy of the system. We define work $W(s)$ as the difference between the mean energy at $s$ and the mean energy at the beginning of the process $\mathcal{E}(0)$, that is, 
\begin{equation}
W(s) \equiv \mathcal{E}(s) - \mathcal{E}(0).
\label{eq: work}
\end{equation}
Another useful quantity is the adiabatic work $W_{\text{ad}}$, which is defined as the work of an infinitely slow process. In this case, the state evolution is given by the adiabatic theorem, Eq. (\ref{eq: adiabatic theorem}), and $\rho(s) = \rho_{\text{ad}}(s) =  | \psi(s) \rangle_{\text{ad}} \langle \psi(s)|_{\text{ad}}$. Mathematically, we have
\begin{equation}
W_{\text{ad}}(s) \equiv \mathcal{E}_{\text{ad}}(s) - \mathcal{E}(0).
\label{eq: quasistatic work}
\end{equation}
The adiabatic work represents the energy cost of the process that causes no transitions among the different energy eigenstates and hence it is inherent to any finite-time process that fulfills the same boundary conditions of the quasistatic one. Thus, the excess work, defined as
\begin{equation}
W_{ex}(s) \equiv W(s) - W_{\text{ad}}(s)\,,
\label{eq: excess work}
\end{equation}
represents the energy pumped into the system due to a finite-time control and essentially is a way to energetically quantify transitions between the energy eigenstates along a quantum process. It strongly depends on the details of the process and is zero in the adiabatic limit.

\subsection{Counterdiabatic driving}
Although adiabatic evolution is achieved only for very slow processes, it can be accelerated in finite-time control through a shortcut to adiabaticity. The most representative STA method is known as counterdiabatic driving, which provides the adiabatic evolution 
$|\psi_{0}(s)\rangle_{\text{ad}}$ of a reference Hamiltonian $H_{0}(s)$ in a finite-time process \cite{demirplak2003adiabatic,demirplak2005assisted,demirplak2008consistency,berry2009transitionless}. To accomplish this, it is necessary to add an extra Hamiltonian $H_{1}(s)$ to $H_{0}(s)$, leading to a new total Hamiltonian $H(s) = H_{0}(s) + H_{1}(s)$. In addition, it is usually required that $H_1(0) = H_1(1) = 0 $, which leads to $H(0) = H_{0}(0)$ and $H(1) = H_{0}(1)$. Therefore, at the end of the process, not only is the time-evolved state equal to the adiabatically evolved $H_{0}$-state, but also the total Hamiltonian is equal to the reference Hamiltonian $H_{0}$. 

Due to the structure of the total Hamiltonian $H(s)$, it will be useful to distinguish its eigenenergies and eigenstates from the original Hamiltonian $H_{0}(s)$. The total Hamiltonian has eigenenergies and eigenstates denoted by $E_{n}(s)$ and $|n(s)\rangle$, respectively. The reference Hamiltonian $H_{0}(s)$ has eigenenergies $E_{n}^{(0)}(s)$ and eigenstates $|n_{0}(s)\rangle$. Hence, subscripts and superscripts $0$ refer to quantities connected with $H_{0}(s)$.

According to Ref.~\cite{berry2009transitionless}, the extra Hamiltonian must have the following form,
\begin{equation}
H_{1}(s) = i \hbar \sum_{m,n,m \neq n} M_{mn}(s)|m_{0}(s)\rangle \langle n_{0}(s)|,
\label{eq: CD_Hamiltonian}
\end{equation}
where
\begin{equation}
\begin{split}
M_{mn}(s) &\equiv \langle m_{0}(s) | \frac{d}{dt} |n_{0} (s) \rangle = \frac{1}{\tau}\langle m_{0}(s) | \frac{d}{ds} |n_{0} (s) \rangle\,,
\end{split}
\end{equation}
to guarantee that the evolution of an initial eigenstate of $H_{0}$ is given by 
\begin{equation}
|\psi(s) \rangle = |\psi_{0}(s)\rangle_{\text{ad}} = e^{i \gamma_n^{(0)} (s)}e^{-i \omega_n^{(0)}(s)}|n_{0}(s) \rangle,
\label{eq: evolution_CD}
\end{equation}
where the phases are given respectively by Eqs.~(\ref{eq: geometric phase}) and 
(\ref{eq: dynamical phase}) with  eigenenergies $E_{n}^{(0)}(s)$ and eigenstates $|n_{0}(s)\rangle$.

\begin{figure}%
    \centering
    {{\includegraphics[width=8cm]{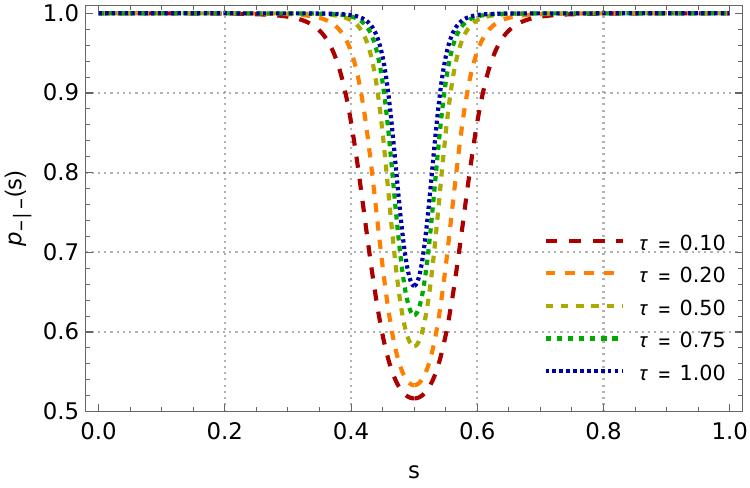} }}%
    \caption{Probability of measuring the instantaneous ground state $|- \rangle$ of the total Hamiltonian $H(t) = H_{0}(t)+H_{1}(t)$ of the Landau-Zener model for different protocol times $\tau$. The Hamiltonian $H_{0}(t)$ is given by Eq.~(\ref{eq: LZ_Hamiltonian}) and $H_{1}(t)$ is given by Eq.~(\ref{eq: CD_LZ}), when counterdiabatic driving is applied during the protocol $B(t)$ given by Eq. (\ref{eq: protocol}). Simulations were performed with $B_{i}/J = -B_{f}/J = -10$ and $J=5$.} 
   \label{fig: Probability_fundamental}%
\end{figure}

It is important to note that counterdiabatic driving only achieves the adiabatic evolution of the reference Hamiltonian $H_{0}(s)$, while the dynamics due to the total Hamiltonian remains diabatic. Thus, while transitions among the energy eigenstates of $H_{0}(s)$ are suppressed, transitions do occur among the eigenstates of the total Hamiltonian $H(s)$, as illustrated in Fig.\ref{fig: Probability_fundamental} for the Landau-Zener model. 

\section{Quantum speed limits for driven systems}
\label{sec: MLbound}

\begin{figure}%
    \centering
    {{\includegraphics[width=8cm]{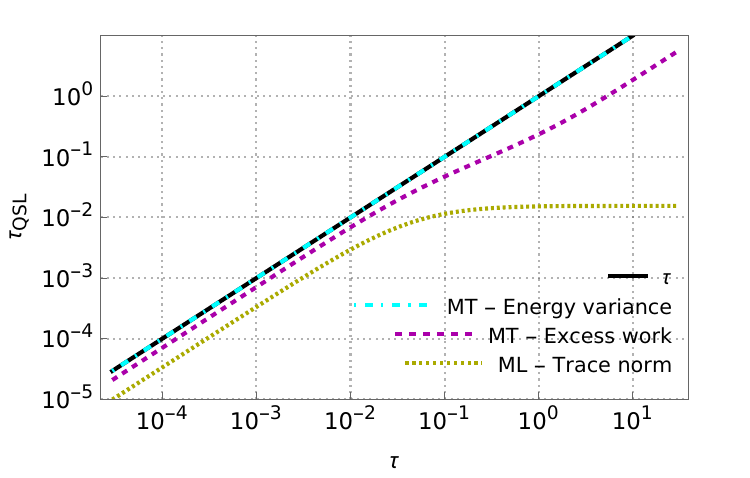} }}%
    \caption{Quantum speed limit in the driven Landau-Zener model as a function of the protocol time $\tau$. In cyan/dashed is $\tau_{QSL}$ given by the Mandelstam-Tamm bound (Eq.(\ref{eq: QSL})). In magenta/dashed is the Mandelstam-Tamm bound given in terms of the excess work (Eq.(\ref{eq: QSL_excess_2})) using Eq.(\ref{eq: excess_work_new}). In yellow/dashed is the Margolus-Levitin bound of reference \cite{deffner2013quantum} given in terms of the trace norm of $\rho(t)H(t)$.  All quantities must be equal or smaller than $\tau$ (solid/black), as can be verified in the figure. The parameters used were $B_{i}/J = -B_{f}/J = 10$ and $J = 5$ with protocol  $B(s) = B_{i} + (B_{f} - B_{i})s$. It is noteworthy that the Mandelstam-Tamm bound is tight in this case.} 
    \label{fig: QSL_Excess_work}%
\end{figure}

Heisenberg's uncertainty principles \cite{heisenberg1927anschaulichen}, whose version involving momentum and position is the most representative example, is at the heart of quantum mechanics. Nevertheless, the uncertainty-like relation between time and energy has challenged scientists for nearly 100 years. This is so because typical uncertainty principles are understood as statements about the impossibility of simultaneous measurements, whereas the uncertainty-like relation between time and energy has the interpretation of the minimal time that a quantum state can evolve given a certain amount of energy. The bound for the minimal time of evolution of a quantum state is set, for time-independent Hamiltonians, by either the Mandelstam-Tamm bound (MT) \cite{mandelstam1945uncertainty} or the Margolus-Levitin (ML) bound \cite{margolus1998maximum}, where the former depends on the variance of the energy, while the latter depends on the mean energy. These bounds are part of a set of similar relations called quantum speed limits, since they impose maximum rates for the evolution of quantum states that are also extended for driven dynamics \cite{deffner2013energy}. In this case, they establish that a faster evolution demands an energy input into the system or a spreading of energy over the energy eigenstates, implying then trade-off relations between the evolution speed and the amount of energy required for it. The Mandelstam-Tamm bound for closed driven dynamics reads \cite{deffner2013energy, deffner2017quantum, braunstein1995dynamics, braunstein1996generalized}, 
\begin{equation}
\tau \geq  \tau_{\text{QSL}} \equiv \frac{\hbar   \mathcal{L}(\rho(t_{i}), \rho(t_{f}))}{ \frac{1}{\tau} \int_{t_{i}}^{t_f} \text{Var}[H(t)] dt  },
\label{eq: QSL}
\end{equation}
where $\text{Var}[H(t)]$ is the energy variance along the driven evolution, $\mathcal{L}(\rho(t_{i}), \rho(t_{f}))$ is the Bures angle between the initial density matrix, $\rho(t_{i})$, and the time-evolved one, $\rho(t_{f})$, after a time interval $\tau=t_{f}-t_{i}$, which is also the duration of the applied time-dependent control. The right-hand side of Eq.(\ref{eq: QSL}) then represents the minimum time that a driven system can evolve from the state $\rho(t_{i})$ to the target state $\rho(t_f)$. For the Landau-Zener model under counterdiabatic driving, the Mandelstam-Tamm bound is tight, as can be seen in Fig. \ref{fig: QSL_Excess_work}.

If the system is initially in the ground state of the Hamiltonian, the distribution of energy among the different eigenstates during the process necessarily involves transitions to excited eigenstates, i.e., transitions throughout the process are necessary and strongly connected with the quantum speed limit. Moreover, faster processes cause more transitions and, consequently, spread more energy. Typically, these transitions can also be assessed using excess work; therefore, we will rewrite Eq. (\ref{eq: QSL}) in terms of excess work. 

\begin{figure}%
    \centering
    {{\includegraphics[width=8cm]{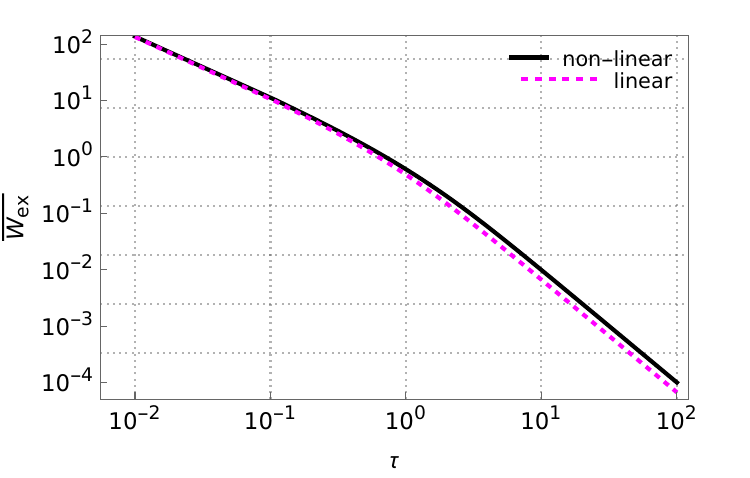} }}%
    \caption{Time average of expression (\ref{eq: excess_work_new}) for the excess work as a function of the driving duration for the protocols given by Eq. (\ref{eq: protocol}) (solid/black) and for the linear protocol, $B(s) = B_{i} + (B_{f} - B_{i})s$ (dashed/magenta). Note the non-zero value of the time-averaged excess work and the decaying  with the protocol time $\tau$. The parameters used were $B_{i}/J = -B_{f}/J = 10$ and $J = 5$.} 
    \label{fig: Excess_work_tau}%
\end{figure}

First, consider a finite-dimensional system, with $N$ non-degenerate energy eigenvalues $(E_1, E_2, E_3,..., E_N)$ organized in ascending order and that $\Delta E(t) \equiv E_N(t) - E_1(t)$ represents the difference between the larger and smaller eigenvalues of $H(s)$. Moreover, consider that the system is initially in the ground state, so that $\mathcal{E}_{\text{ad}}(t) = E_{1}(t)$. So, using the Bhatia-Davis inequality \cite{bhatia2000better}, we can write the energy variance as
\begin{equation}
\begin{split}
\text{Var}[H(t)]^2 & \leq [E_{N}(t) - \mathcal{E}(t)][\mathcal{E}(t) - E_{1}(t) ]\\
& = [\Delta E(s) - W_{ex}(t)]W_{ex}(t) \\
& \leq \Delta E(t)  W_{ex}(t).
\end{split}
\end{equation}
Substituting the above inequality into Eq.(\ref{eq: QSL}), we obtain
\begin{equation}
\tau \geq \frac{\hbar   \mathcal{L}(\rho(t_{i}), \rho(t_{f}))}{ \frac{1}{\tau} \int_{t_{i}}^{t_f} \sqrt{\Delta E(t)  W_{ex}(t)} dt  }.
\label{eq: QSL_excess}
\end{equation}
In addition, using the Cauchy-Schwarz inequality \cite{clapham2014concise}, we can write
\begin{equation}
 \int_{t_{i}}^{t_f} \sqrt{\Delta E(t)  W_{ex}(t)} dt \leq \sqrt{\int_{t_i}^{t_f}\Delta E(t) dt \int_{t_i}^{t_f} W_{ex}(t) dt} 
\end{equation}
and, consequently,
\begin{equation}
\tau \geq \frac{\hbar   \mathcal{L}(\rho(t_{i}), \rho(t_{f}))}{ \sqrt{\overline{\Delta E}} \sqrt{\overline{W_{ex}}} },
\label{eq: QSL_excess_2}
\end{equation}
where we defined the time-averaged quantities,
\begin{equation}
\overline{ \Delta E}  \equiv  \frac{1}{\tau}\int_{0}^{\tau} \Delta E(t) dt
\end{equation}
and,
\begin{equation}
\overline{ W_{ex}}  \equiv \frac{1}{\tau}\int_{0}^{\tau} W_{ex}(t) dt.
\end{equation}
Equation (\ref{eq: QSL_excess_2}), obtained from the Mandelstam-Tamm bound (Eq.(\ref{eq: QSL})), gives a quantum speed limit dependent on two quantities: the time-averaged maximum energy gap and the time-averaged excess work. Although it represents a less tight bound than the Mandelstam-Tamm bound, the example of the driven Landau-Zener model shows that it still gives a tighter bound than the one given by the Margolus-Levitin bound for closed quantum systems \cite{deffner2013quantum}, see Fig. \ref{fig: QSL_Excess_work}. Although the time-averaged maximum energy gap is intrinsic to the control of the quantum system through the Hamiltonian and protocol, the time-averaged work strongly depends on the dynamics, energetically quantifying the transitions induced by the finite-time protocol. Additionally, $\overline{W_{ex}}$ cannot be zero and increases for faster protocols, as illustrated in Fig. \ref{fig: Excess_work_tau}, and thus represents an energy input to speed up the quantum process. Therefore, Eq.(\ref{eq: QSL_excess_2}) shows a trade-off relation between time and the time-averaged excess work. We propose that such a quantity gives valuable information about the energetics of the system even in counterdiabatic driving.
 
\section{Time-averaged excess work in counterdiabatic driving}
\label{sec: time-averaged_work_in_cd}
The counterdiabatic method suppresses transitions between different eigenstates of $H_{0}(t)$ at all instants of the protocol, for arbitrarily fast processes. Considering then trade-off relations just discussed provided by quantum speed limits, one could expect that this speed-up of the quantum control would be accompanied by a higher energetic cost. It seems natural then to use the time-averaged excess work as a quantifier of this cost, since counterdiabatic driving is another example of time-dependent Hamiltonian control. However, the application of excess work for counterdiabatic driving is not as straightforward as one might expect. Work, as defined in Eq.~(\ref{eq: work}), is indeed equal to its adiabatic value (\ref{eq: quasistatic work}) when counterdiabatic driving is applied. To better see this, we can start calculating the mean energy at any instant of the process. It reads as follows
\begin{equation}
\begin{split}
\mathcal{E}(s)& = \text{Tr}\{\left[H_{0}(s) + H_{1}(t)\right]\rho_{\text{ad}}^{(0)}(s)  \} \\
& = \text{Tr}\{H_{0}(s)\rho_{\text{ad}}^{(0)}(s) \} +  \text{Tr}\{H_{1}(s)\rho_{\text{ad}}^{(0)}(s) \} \\
& =  \text{Tr}\{H_{0}(s)\rho_{\text{ad}}^{(0)}(s) \} 
\end{split}
\end{equation}
since
\begin{equation}
 \text{Tr}\{H_{1}(s)\rho_{\text{ad}}(s)\} = 0  
\end{equation}
using Eqs. (\ref{eq: CD_Hamiltonian}), (\ref{eq: evolution_CD}) and $\rho_{\text{ad}}^{(0)}(t) = |\psi_{0}\rangle_{\text{ad}} \langle \psi_{0}(t)|_{\text{ad}} $. Hence, $H_{1}(t)$ does not contribute to the mean energy and the work in counterdiabatic driving is equal to
\begin{equation}
\begin{split}
W(s) &=  \text{Tr}\{H_{0}(s)\rho_{ad}(s) \}-  \text{Tr}\{H_{0}(0)\rho(0)\}\\
& = W_{\text{ad}}^{(0)}(s).
\end{split}
\end{equation} 
The upper index $(0)$ indicates that $W^{(0)}_{\text{ad}}(s)$ is obtained from $H_{0}$ only. 

In the situation where the initial state is an eigenstate of $H_{0}(0)$, for example $|n_{0}(0)\rangle$, the work reads
\begin{equation}
W(s) = E_{n}^{(0)}(s) - E_{n}^{(0)}(0).
\label{eq: work_cd}
\end{equation}
and the adiabatic contribution along the same process can be written as
\begin{equation}
W_{\text{ad}}(s) = E_{n}(s) -E_{n}^{(0)}(0) . 
\label{eq: work_ad}
\end{equation}
as demonstrated in App.~\ref{app: work} (Eq.(\ref{eq: adiabatic_work})). Hence, the excess work can be written as
\begin{equation}
W_{ex}(s) =  E_{n}^{(0)}(s) - E_{n}(0).
\label{eq: excess_work_pure}
\end{equation}
However, since in the adiabatic limit we have 
\begin{equation}
\begin{split}
\lim_{\tau \rightarrow \infty} & H_{1}(s) = \\
&\lim_{\tau \rightarrow \infty}\frac{i \hbar}{\tau} \sum_{m,n,m \neq n} \langle m_{0}(s) | \frac{d}{ds} |n_{0} (s) \rangle|m_{0}(s)\rangle \langle n_{0}(s)|\\
&= 0
\end{split}
\label{eq: adiabatic_limit_H1}
\end{equation}
the eigenvalues of $H(s)$ and $H_{0}(s)$ coincide in such limit, that is, $\lim_{\tau \rightarrow \infty}E_{n}(s) = E_{n}^{(0)}(s)$ and, 
\begin{equation}
W_{ex}(s) = 0 \rightarrow \overline{W}_{ex} = 0. 
\end{equation}

Therefore, the excess work is zero for any $s$, which is a surprising result if we consider that it energetically quantifies the transitions between the eigenstates of the total Hamiltonian. On the other hand, as we have seen in the previous section by using the Mandelstam-Tamm bound, the time-averaged excess work should be different from zero. To reconcile these two results, we propose to change the perspective about the limit (\ref{eq: adiabatic_limit_H1}), which, as we will show next, leads to a different adiabatic work and, consequently, to a non-vanishing excess work.

\begin{figure}
    \centering
    {{\includegraphics[width=8cm]{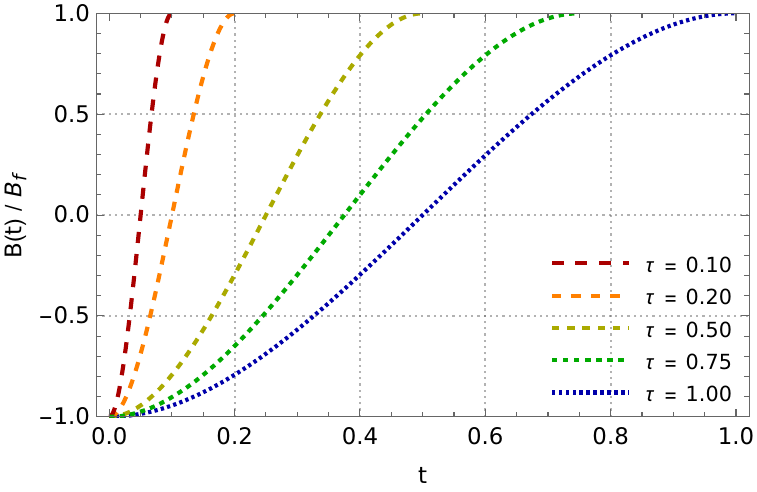} }}%
    \caption{Protocol $B(t)$ of Eq.~(\ref{eq: protocol}) for different durations. As we increase $\tau$, the relative change of $B$ (which is independent of $\tau$) takes place at smaller rates. The values used for initial and final values were $B_{i}=-50$ and $B_{f}=50$, respectively.} 
    \label{fig: Protocols_AT}
\end{figure}

\section{The double role of the adiabatic limit}
\label{sec: AdiabaticLimit}

\begin{figure}%
    \centering
    {{\includegraphics[width=8cm]{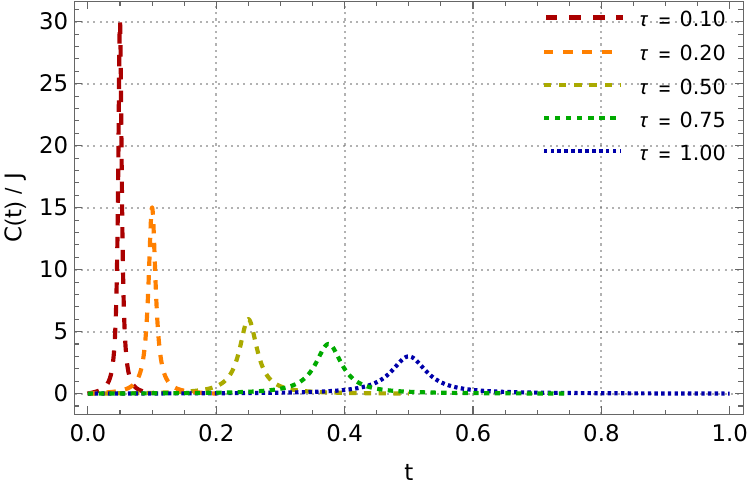} }}%
    \caption{Variation of the amplitude $C(s)$, given by Eq.(\ref{eq: CD_LZ}), of the counterdiabatic Hamiltonian of the Landau-Zener model. The different curves represent processes with different durations. As $\tau$ increases, the intensity of $C(t)$ becomes smaller.} 
    \label{fig: CD_Protocol_AT_TypeI}%
\end{figure}

Let us consider two time-dependent drivings applied in the same system, with the same boundary conditions, that is, starting from the same initial value of the control parameter and ending at the same final value of it. If the only difference between these two drivings is their duration, we would expect a higher cost to perform the faster one. In the presence of the counterdiabatic Hamiltonian $H_{1}$, despite what has been just discussed for the excess work, the situation is not much different from this, as Figs.~\ref{fig: Probability_fundamental} and \ref{fig: CD_Protocol_AT_TypeI} show for the Landau-Zener model (which we shall discuss in more detail next). Figure \ref{fig: Probability_fundamental} shows that the probability of remaining in the ground state of the total Hamiltonian has a deep that decreases monotonically with the increasing duration $\tau$ of the driving depicted in Fig. \ref{fig: Protocols_AT}. In other words, transitions occur more intensively between the eigenstates of the total Hamiltonian as the driving becomes faster. Figure \ref{fig: CD_Protocol_AT_TypeI} shows that the intensity of the counterdiabatic Hamiltonian $H_{1}$ also increases as the driving speeds up, which is another signature of higher cost. 

A related and interesting point to note is that the instantaneous eigenvalues of the total Hamiltonian depend on the driving duration. Hence, two drivings differing only by their $\tau$'s lead to different eigenvalues of the total Hamiltonian simply because their corresponding $\tau$'s provide different intensities of $H_{1}$, as shown in Eq.~(\ref{eq: CD_Hamiltonian}). This is a feature that distinguishes the total Hamiltonian generated by the counterdiabatic method from the usual time-dependent Hamiltonians in which the intensity of the driving is independent of its speed. In the standard cases, drivings with different intensities but identical protocols for the variation of the control parameter (differing perhaps only by their duration, as depicted in Fig. \ref{fig: Protocols_AT}) provide \emph{different} instantaneous eigenenergies that play a crucial role in the definition of work.

In the case of counterdiabatic driving, we then propose to take as the eigenenergies of the total Hamiltonian that enter in the adiabatic work the instantaneous eigenvalues of $H(t)$, which will be functions of $\lambda(s)$, $\lambda(s)$ being the driving protocol for the control parameter $\lambda$, but also functions of $d\lambda(s)/ds$ and $\tau$. The protocol time $\tau$ is then kept fixed, identical to the original driving duration and understood as providing the intensity of the counterdiabatic contribution $H_{1}$. In other words, from now on we will denote $H_{1}$ by
\begin{equation}
H_{1}(s) = \frac{i \hbar}{\tau_{d}} \sum_{m,n,m \neq n} \langle m_{0}(s) | \frac{d}{ds} |n_{0} (s) \rangle|m_{0}(s)\rangle \langle n_{0}(s)|.
\label{eq: counterdiabatic}
\end{equation}
where $\tau_{d}$ stands for the driving duration of the original finite-time process. Consequently, $\lim_{\tau\to\infty} H_{1}$ no longer vanishes. In addition, 
\begin{equation}
\lim_{\tau \rightarrow \infty}E_{n}(s) \neq E_{n}^{(0)}(s)\,,
\end{equation}
for all $s$ except $s=0$ and $s=1$ where $H_{1}$ vanishes anyway due to the chosen boundary conditions. This tells us that the instantaneous eigenvalues of $H(t)$, with the original driving protocol, are the parametrically changing eigenenergies of the adiabatic limit, as in the standard time-dependent Hamiltonian drivings.

Next, we will check the consequences of this prescription more concretely with the example of the driven Landau-Zener Hamiltonian subjected to counterdiabatic driving.
But before, note that parameterization $s = t/\tau$ was essential in writing Eq.(\ref{eq: counterdiabatic}), as it clearly separates the counterdiabatic terms into two parts: one that determines the strength of the control, $ds/dt = 1/\tau_d$, and one that sets the shape of the protocol, $\langle m_{0}(s) | \frac{d}{ds} |n_{0} (s) \rangle$. In the adiabatic limit, the proposed procedure consists in keeping constant the part that determines the strength, i.e., $ds/dt = 1/\tau_d$. Even when other parameterizations of the driving protocol $\lambda(t)$ are used, we stress that our proposal is to keep $ds/dt = 1/\tau = 1/\tau_{d}$ constant in the adiabatic limit.

\subsection{Example: The Landau-Zener model}
\label{sec: LZ-Model}
The Landau-Zener Hamiltonian can be written as
\begin{equation}
H_{0}(t) = B(t) \sigma^z + J \sigma^x
\label{eq: LZ_Hamiltonian}
\end{equation}
where $\sigma^i$, with $i = x,y,z$, represents the Pauli matrices and $B(t)$ the control parameter. Its instantaneous eigenenergies are given by
\begin{equation}
E_{\pm}^{(0)}(s) =\pm \sqrt{B(s)^2 + J^2}
\label{eq: Eigenenergies_H0}
\end{equation}
and the corresponding eigenstates read
\begin{equation}
\begin{split}
| E_{+}^{(0)} \rangle &= \cos \theta | \uparrow \rangle + \sin \theta | \downarrow \rangle \\
| E_{-}^{(0)} \rangle &= - \sin \theta | \uparrow \rangle + \cos \theta | \downarrow \rangle 
\end{split}
\label{eq: eigenstatesH0}
\end{equation}
where $\sigma^z | \uparrow \rangle =   | \uparrow \rangle $ and $\sigma^z | \downarrow \rangle = -  | \downarrow \rangle $ and
\begin{equation}
\theta \equiv \frac{1}{2} \tan^{-1}\left(\frac{J}{B}\right).
\end{equation}

The counterdiabatic Hamiltonian $H_{1}(t)$ necessary to achieve the adiabatic evolution of the ground state of Hamiltonian (\ref{eq: LZ_Hamiltonian}) in finite time is written as
\begin{equation}
H_{1}(s) = \frac{\hbar J \,\dot{B}(s)}{2 \tau_{d} \left[E_{+}^{(0)}(B(s))\right]^2}\, \sigma^y = C(s) \sigma^{y},
\end{equation}
where, $\dot{B}(s)=dB/ds$ and
\begin{equation}
C(s) \equiv \frac{\hbar J \,\dot{B}(s)}{2 \tau_{d} \left[E_{+}^{(0)}(B(s))\right]^2}\,. 
\label{eq: CD_LZ}
\end{equation}

Although not usual in applications of the counterdiabatic method, it is important for our purposes to also know the spectrum of the total Hamiltonian $H(s) = H_{0}(s) + H_{1}(s)$,
\begin{equation}
E_{\pm}(s) = \pm \sqrt{B(s)^2 + J^2 + C(s)^2}
\label{eq: Eigenenergies_Tot}
\end{equation}
and its corresponding eigenstates, 
\begin{equation} 
\begin{split}
|+ \rangle  & = \cos \theta_{c} e^{i\mu } |\uparrow \rangle + \sin \theta_{c} |\downarrow\rangle\\
  |-\rangle & = - \sin \theta_{c} e^{i\mu } |\uparrow \rangle + \cos \theta_{c} |\downarrow\rangle ,
\label{eq: eigenstatesH}
\end{split}
\end{equation}
where
\begin{equation}
\theta_c = \frac{1}{2}\tan^{-1}\left(\frac{\sqrt{C^2 + J^2}}{B}\right)
\end{equation}
and
\begin{equation}
\mu = \tan^{-1} \left( \frac{C}{J}\right).
\end{equation}

\begin{figure}%
    \begin{center}
    \centering
    {{\includegraphics[width=8cm]{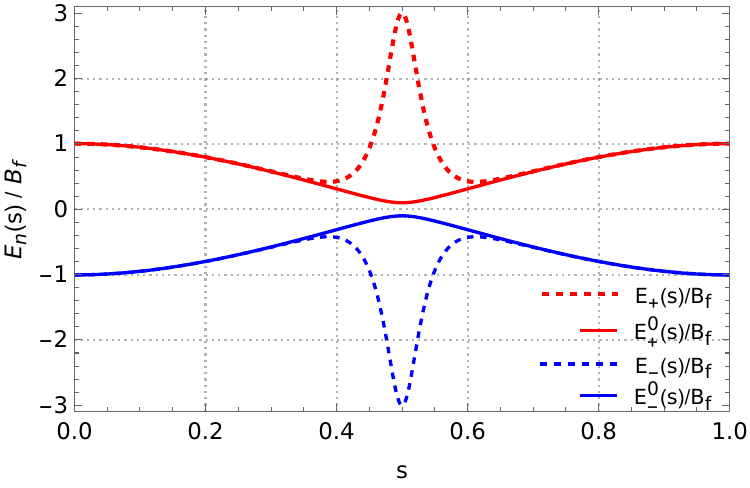} }}%
    \end{center}
    \caption{Variation of the instantaneous eigenenergies (\ref{eq: Eigenenergies_H0}) of the Hamiltonian $H_{0}(t)$ and (\ref{eq: Eigenenergies_Tot}) of $H(t)$ during the protocol given by Eq. (\ref{eq: protocol}) with $\tau = 0.1$. At the avoided crossing of $H_{0}(t)$, $t = 0.05$, the spectrum of $H_{1}(t)$ has its maximal gap. The remaining parameters were set as $J=5$ and $B_{i}=-50 = -B_{f}$.} 
    \label{fig: Eigenenergies}%
\end{figure}

The parametric change of the instantaneous eigenenergies (\ref{eq: Eigenenergies_Tot}) of the total Hamiltonian and (\ref{eq: Eigenenergies_H0}) of $H_{0}(s)$ are represented in Fig.~\ref{fig: Eigenenergies} using the protocol \cite{campos2018error}
\begin{equation}
    B(t) = B_i + (B_f - B_i)\left( 3 s^{2} - 2 s^{3} \right)\,.
\label{eq: protocol}
\end{equation}
It shows that the maximal gap of $H(s)$ occurs exactly at the minimum gap of $H_{0}(s)$.

Using Eqs.~(\ref{eq: work_ad}), (\ref{eq: Eigenenergies_Tot}) and the prescription discussed previously, namely, to keep $\tau_{d}$ \emph{finite} in the adiabatic limit of $H_{1}(s)$, we find that the adiabatic work for an intermediate time $s$ of the control is equal to 
\begin{equation}
W_{{\text{ad}}}(s) = - \sqrt{B(s)^2 + J^2 + C(s)^2} + \sqrt{B(0)^2 + J^2}
\end{equation}
and, therefore, from Eq.~(\ref{eq: excess_work_pure}), we obtain
\begin{equation}
\begin{split}
    W_{ex}(s) &= - \sqrt{B(s)^2 + J^2} +  \sqrt{B(s)^2 + J^2 + C(s)^2}\\
    &=- \sqrt{B(s)^2 + J^2} \\
    &+   \sqrt{B(s)^2 + J^2 + \left[\frac{1}{2\tau_{d}} \frac{\hbar J \dot{B}(s)}{\left[B(s)^2 + J^2\right]}\right]^2}\,,
\end{split}
\label{eq: excess_work_new}
\end{equation}
which does not vanish, in general, for intermediate times $s$ and hence leads to a non-vanishing time-averaged excess work.

\begin{figure}%
    \centering
    {{\includegraphics[width=8cm]{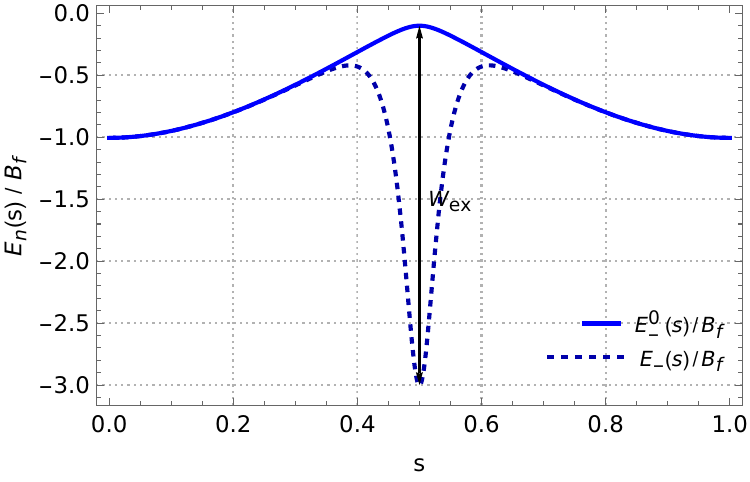} }}%
    \caption{Energies of the ground states of the total Hamiltonian (blue dashed) and of the reference Hamiltonian (blue solid) of the Landau-Zener model for $\tau_{d} = 0.1$. The protocol (\ref{eq: protocol}) was used with $B_{i}/J = -B_{f}/J = -10$ and $J=5$. The difference between the two eigenenergies at each $s$ can be taken as the excess work. See Sec.~\ref{sec: AdiabaticLimit} for the details. } 
    \label{fig: Interpretation_excess_work}%
\end{figure}

This expression for excess work in counterdiabatic driving is simply the difference between the relevant eigenenergies of the reference Hamiltonian and the total Hamiltonian as represented in Fig.~\ref{fig: Interpretation_excess_work}. Furthermore, its time average is sensitive both to the driving speed and the form of the protocol, as shown in Fig.~\ref{fig: Excess_work_tau}. These are two natural features to be expected from an energy-cost quantifier.
 
\section{Discussion}
\label{sec: Discussion}
\subsection{Expected features of an energy cost}

Quantifying the energetic cost of shortcuts to adiabaticity has been an elusive task, especially with regard to its physical interpretation as consumption of thermodynamic resources and valid for general time-dependent Hamiltonians. Next, we enumerate the expected features of an energetic cost for thermally isolated quantum systems:

\begin{itemize}
\item  \emph{Decaying with protocol time}. As the quasistatic limit is approached, observable quantities are expected to converge to their quasistatic value. Hence, we expect a decay of the energetic cost with the driving time. Figure~\ref{fig: Excess_work_tau} shows that the time-averaged excess work fulfills this expectation. 

\item \emph{Dependence on the protocol}. The evolution of the quantum state is essentially related to the form of the driving control. Therefore, we expect that the energetic cost also depends on it \cite{abah2019energetic}. The time-averaged excess work also meets this requirement, as illustrated in Fig.~\ref{fig: Excess_work_tau}. 

\item \emph{Connection to the quantum speed limit}.
The quantum speed limit (QSL) for driven systems imposes the minimum time that a quantum state can evolve to a target state. To this end, it also points out that there must be a consumption of resources \cite{deffner2010generalized}.  A quantum state can change more quickly if energy is injected into the system or if it is more distributed over the energy eigenstates \cite{deffner2013energy}. Considering the connection between QSL and resource consumption, all the energetic costs proposed in Refs.~ \cite{santos2015superadiabatic,zheng2016cost, campbell2017trade, funo2017universal}, obey trade-off relations between the speed of the quantum evolution and the respective energetic cost. The time-averaged excess work we propose in this work was directly constructed from a trade-off relation as such. 
\end{itemize}

\section{Conclusion}
\label{sec: Conclusion}
In this paper, we have shown that the Mandelstam-Tamm bound can be written in terms of the excess work. Nevertheless, at first glance, application of this energy-cost quantifier to counterdiabatic driving appears to fail. To redeem this, we notice that this shortcut method has a specificity: the protocol time has two very distinct roles, namely, it controls the intensity of the counterdiabatic Hamiltonian and, at the same time, gives the rate at which the driving is performed. Two finite-time drivings with different speeds then lead to different instantaneous eigenvalues of the total Hamiltonian, which can be used to quantify the energy cost at different driving rates if the original intensity of the counterdiabatic Hamiltonian is kept. Additionally, we have shown that this does not cause any trouble to the quantum speed limit from which our proposal was motivated.
Thus, by understanding the two different roles of the driving time $\tau$, we can reintroduce excess work as an energy-cost quantifier in counterdiabatic driving as the difference between the instantaneous eigenenergies of the total and reference Hamiltonians.

In addition, we have shown that the time-averaged work has important characteristics that one could expect from a cost quantifier: the decay with protocol time, the dependence on the protocol, and the connection with a trade-off relation between speed and cost. Hence, the time-averaged work fulfills the minimum expectations of an energy cost from a theoretical perspective.  It remains open for further investigations whether this proposal will be useful as a good and physically meaningful energetic cost quantifier for practical purposes. 

\section*{Acknowledgements \label{sec:acknowledgements}}

LPK and MVSB acknowledge financial support by FAPESP (Funda\c{c}\~ao de Amparo \`a Pesquisa do Estado de S\~ao Paulo), Grants n.o 2022/04627-7 and n.o 2022/15453-0. MVSB also acknowledges financial support by CNPq (Conselho Nacional de Desenvolvimento Cient\'ifico e Tecnol\'ogico), Grant n.o 304120/2022-7. 

\bibliography{Ref.bib}
\appendix
\section{Exact expression for work}
\label{app: work}
The work performed on a generic time-dependent Hamiltonian system (with nondegenerate spectrum) up to $s$, such that $0 \leq s < 1$, can be written also in terms of the transition probabilities between the eigenstates of the total Hamiltonian as  
\begin{eqnarray}
    W(s) &=& \mathrm{Tr}\{ H(s) \rho(s)\} - \mathrm{Tr}\{ H(0) \rho(0)\}\nonumber \\
    &=& \sum_{m,n}p_{n} p_{m|n}(s)\left[ E_{m}(s) - E_{n}(0)\right]\,,
\end{eqnarray}
where $\rho(s)$ denotes the density matrix of the time-dependent system at $s$ (assumed to commute with $H(t)$ at $s=0$), $p_{n}$ are the initial statistical weights of the eigenstates of $H(0)$ and $p_{m|n}(s)$ are the transition probabilities from the eigenstate $n$ to $m$ at $s$. Considering that initially the system is in the $l$-th energy eigenstate of the the total Hamiltonian, then $p_{n} = \delta_{l,n}$, and we have
\begin{equation}
W(s) =  \sum_{m} p_{m|l}(s)\left[ E_{m}(s) - E_{l}(0)\right].
\end{equation}
In the adiabatic limit there are no transitions so, $p_{m|l}(s) \rightarrow \delta_{m,l}$, and the adiabatic work reads, 
\begin{equation}
W_{\text{ad}}(s) =  E_{n}(s) - E_{n}(0).
\label{eq: adiabatic_work}
\end{equation}

\end{document}